\documentclass[prd, nofootinbib, aps, twocolumn, superscriptaddress,
showpacs, preprintnumbers, amsmath, amssymb, floatfix]{revtex4}
\usepackage{graphicx,epsfig}
\newcommand{\nc}{\newcommand}
\nc{\hef}{\ensuremath{^4\mathrm{He}}}
\nc{\het}{\ensuremath{^3\mathrm{He}}}
\nc{\lisx}{\ensuremath{^6\mathrm{Li}}}
\nc{\lisv}{\ensuremath{^7\mathrm{Li}}}
\nc{\bes}{\ensuremath{^7\mathrm{Be}}}
\nc{\beet}{\ensuremath{^8\mathrm{Be}}}
\nc{\ben}{\ensuremath{^9\mathrm{Be}}}
\nc{\dm}{{\rm D}}
\nc{\hefm}{{\rm ^4He}}
\nc{\hetm}{{\rm ^3He}}
\nc{\lisxm}{{\rm ^6Li}}
\nc{\lisvm}{{\rm ^7Li}}
\nc{\besm}{{\rm ^7Be}}
\nc{\beetm}{{\rm ^8Be}}
\nc{\benm}{{\rm ^9Be}}
\nc{\bs}{(N$X^-$)}
\nc{\xm}{$X^-$}
\nc{\xp}{$X^+$}
\nc{\xz}{$X^0$}
\nc{\bex}{(\bes\xm)}
\nc{\bexm}{(\besm X^-)}
\nc{\px}{($p$\xm)}
\nc{\Ox}{\ensuremath{\mathrm{O}}}
\nc{\Fe}{\ensuremath{\mathrm{Fe}}}
\nc{\Hyd}{\ensuremath{\mathrm{H}}}
\nc{\Be}{\ensuremath{\mathrm{Be}}}
\nc{\tauX}{\ensuremath{\tau_{X^-}}}
\nc{\YX}{\ensuremath{Y_{X^-}}}
\nc{\YXdec}{\ensuremath{Y^{\mathrm{dec}}_{X^-}}}
\nc{\Ysldec}{\ensuremath{Y^{\mathrm{dec}}_{{\widetilde{l}_1}}}}
\nc{\Ysldecm}{\ensuremath{Y^{\mathrm{dec}}_{{\widetilde{l}_1^-}}}}
\newcommand{\Order}{{\cal O}}   
\newcommand{\GeV}{\mathrm{GeV}}
\newcommand{\TeV}{\mathrm{TeV}}
\newcommand{\Mpc}{\mathrm{Mpc}}
\newcommand{\km}{\mathrm{km}}

\newcommand{\seconds}{\mathrm{s}}

\newcommand{\gravitino}{{\widetilde{G}}}
\newcommand{\axino}{{\widetilde{a}}}
\newcommand{\maxino}{m_{\axino}}

\newcommand{\slepton}{{\tilde{l}_1}}
\newcommand{\stauone}{{\widetilde{\tau}_1}}
\newcommand{\stau}{{\widetilde{\tau}}}
\newcommand{\mstau}{m_{\widetilde{\tau}}}

\newcommand{\stauR}{\ensuremath{{\widetilde{\tau}_{\mathrm{R}}}}}
\newcommand{\selectronR}{\ensuremath{{\widetilde{e}_{\mathrm{R}}}}}
\newcommand{\smuonR}{\ensuremath{{\widetilde{\mu}_{\mathrm{R}}}}}

\newcommand{\sel}{{\widetilde{e}}}
\newcommand{\smu}{{\widetilde{\mu}}}

\newcommand{\st}{{\tilde{\tau}_1}}
\newcommand{\bino}{{\widetilde B}}

\newcommand{\mbino}{m_{\widetilde{B}}}
\newcommand{\neutralino}{{\widetilde \chi}^{0}_{1}}

\newcommand{\cquark}{\ensuremath{\mathrm{c}}}
\newcommand{\anticquark}{\ensuremath{\bar{\mathrm{c}}}}
\newcommand{\bquark}{\ensuremath{\mathrm{b}}}
\newcommand{\antibquark}{\ensuremath{\bar{\mathrm{b}}}}
\newcommand{\mZ}{\ensuremath{M_{\mathrm{Z}}}}

\newcommand{\quark}{\mathrm{q}}
\newcommand{\antiquark}{\bar{\mathrm{q}}}
\newcommand{\PQ}{\mathrm{PQ}}

\newcommand{\NLSP}{\mathrm{NLSP}}

\newcommand{\NTP}{\mathrm{NTP}}
\newcommand{\TP}{\mathrm{TP}}

\newcommand{\CDM}{\mathrm{dm}}

\newcommand{\EM}{\mathrm{em}}
\newcommand{\HAD}{\mathrm{had}}

\newcommand{\OmegaDM}{\Omega_{\mathrm{dm}}}

\newcommand{\Reheating}{\mathrm{R}}
\newcommand{\TR}{T_{\Reheating}}

\newcommand{\Color}{\mathrm{c}}
\newcommand{\Weak}{\mathrm{L}}
\newcommand{\Hypercharge}{\mathrm{Y}}
\newcommand{\Lisix}{{}^6 \mathrm{Li}}
\newcommand{\Hefour}{{}^4 \mathrm{He}}

\newcommand{\decoupling}{\mathrm{dec}}

\newcommand{\deuterium}{\mathrm{D}}

\newcommand{\SUSY}{\mathrm{SUSY}}
\newcommand{\tot}{\mathrm{tot}}

\newcommand{\be}{\begin{equation}}
\newcommand{\ee}{\end{equation}}
\newcommand{\bea}{\begin{eqnarray}}
\newcommand{\eea}{\end{eqnarray}}
\newcommand{\benn}{\begin{displaymath}}
\newcommand{\eenn}{\end{displaymath}}
\newcommand{\beann}{\begin{eqnarray*}}
\newcommand{\eeann}{\end{eqnarray*}}
\begin{document}
%
%
\preprint{ZU--TH 13/09, MPP--2009--155}
%
%
%
\title{Late Energy Injection and Cosmological Constraints\\
  in Axino Dark Matter Scenarios}
\author{Ayres Freitas}
\email{afreitas@pitt.edu}
\affiliation{Department of Physics and Astronomy, 
University of Pittsburgh, 
PA 15260, USA}
\author{Frank Daniel Steffen}
\email{steffen@mppmu.mpg.de}
\affiliation{Max-Planck-Institut f\"ur Physik, 
F\"ohringer Ring 6,
D--80805 Munich, Germany}
\author{Nurhana Tajuddin} 
\email{nurhana@physik.uzh.ch}
\affiliation{Institut f\"ur Theoretische Physik, 
Universit\"at Z\"urich, 
Winterthurerstrasse 190, 
CH--8057~Z\"urich, Switzerland}
\author{Daniel Wyler} 
\email{wyler@physik.unizh.ch}
\affiliation{Institut f\"ur Theoretische Physik, 
Universit\"at Z\"urich, 
Winterthurerstrasse 190, 
CH--8057~Z\"urich, Switzerland}
%
%
\begin{abstract}
  Taking into account effects of late energy injection, we
  examine big bang nucleosynthesis (BBN) constraints on axino dark
  matter scenarios with long-lived charged sleptons.
  We calculate 4-body slepton decays into the axino, a lepton, and a
  quark-antiquark pair since they govern late hadronic energy
  injection and associated BBN constraints.
  For supersymmetric hadronic axion models,
  we present the obtained hadronic BBN constraints and show that they
  can be more restrictive than the ones associated with catalyzed BBN
  via slepton-bound-state formation.
  From the BBN constraints on hadronic and electromagnetic energy
  release, we find new upper limits on the Peccei--Quinn scale.
\end{abstract}
\pacs{98.80.Cq, 95.35.+d, 12.60.Jv, 95.30.Cq}
%
%
\maketitle
%
\section{Introduction}

It is widely believed that the Standard Model of particle physics is
not complete and that new physics is required for a compelling
theoretical description of nature. The Large Hadron Collider (LHC) may
give us a first opportunity to test models of new physics directly.

In addition to collider experiments, the early Universe offers many
ways to probe new physics. In particular, the observed dark matter
density and the framework of big bang nucleosynthesis (BBN) impose
valuable restrictions on the parameters of new physics models.

Supersymmetry (SUSY) and the Peccei--Quinn (PQ) symmetry are
particularly well-motivated extensions of the Standard Model each of
which comes with a compelling dark matter candidate. If both of these
extensions are realized simultaneously, significant contributions to
the cold dark matter density $\Omega_{\CDM}$ can reside in the axion
and/or the lightest supersymmetric particle (LSP).

An attractive realization of this scenario with conserved R-parity has
the axino $\axino$---the fermionic partner of the axion---as the LSP
and a charged slepton $\slepton$ as the next-to-lightest
supersymmetric particle (NLSP). In fact, the axino is an extremely
weakly interacting particle and a promising dark matter candidate
beyond the minimal supersymmetric Standard Model
(MSSM)~\cite{Bonometto:1993fx,Covi:1999ty,Covi:2001nw,Covi:2004rb,Brandenburg:2004du,Steffen:2008qp,Baer:2008yd,Freitas:2009fb}.
Since the axino LSP allows for a long-lived charged slepton---such as
the lighter stau $\stauone$---which should be easy to discover at the
LHC, those scenarios are appealing not only from a theoretical point
of view but also from a phenomenological one.

The requirement of avoiding overclosure of the Universe from an overly
efficient production of axino dark matter allows one to derive upper
limits on the reheating
temperature~\cite{Covi:2001nw,Brandenburg:2004du,Choi:2007rh,Kawasaki:2007mk,Baer:2008yd,Freitas:2009fb,Baer:2009ms}.
In our recent Letter~\cite{Freitas:2009fb}, we have shown how those
constraints become more restrictive in the above scenarios once
catalyzed BBN (CBBN) and the associated production of primordial
lithium--6 and
beryllium--9~\cite{Pospelov:2006sc,Pospelov:2007js,Pospelov:2008ta}
are considered.  Moreover, while axion physics provides us with a
lower limit~\cite{Sikivie:2006ni,Raffelt:2006cw,Amsler:2008zz}
\begin{equation}
  f_a\gtrsim 6\times 10^8\,\GeV
  \, ,
\label{Eq:f_a_axion}
\end{equation}
we found that CBBN also imposes upper limits on $f_a$ that depend on
the slepton mass~\cite{Freitas:2009fb}.

In this Letter we further develop the BBN constraints on axino LSP
scenarios with long-lived charged sleptons by considering hadronic and
electromagnetic energy injection.
If the $\slepton$ is relatively long-lived, decay modes with hadrons
in the final state can affect BBN and are thus strongly restricted. We
analyze the most important contribution of this kind: the hadronic
4-body decay of $\slepton$ into the associated lepton, the axino, and
a quark-antiquark pair.
We also include the electromagnetic energy injection, which is derived
from the 2-body decay of the $\slepton$ into the the associated lepton
and the axino.
In fact, the BBN constraints on hadronic and electromagnetic energy
release allow us to derive new upper limits on $f_a$ which appear in
addition to those imposed by CBBN.

\section{Particle physics setting}
\label{Sec:Particle_Physics_Setting}

We consider SUSY hadronic (or KSVZ~\cite{Kim:1979if,Shifman:1979if})
axion models~\cite{Kim:1983ia} in which the interaction of the axion
multiplet $\Phi$ with the heavy KSVZ quark multiplets $Q_1$ and $Q_2$
is described by the superpotential
\begin{equation}
        W_{\PQ}=y\Phi Q_1 Q_2 
\label{Eq:SuperpotentialPQ}
\end{equation} 
with the Yukawa coupling $y$ and the quantum numbers given in
Table~\ref{Tab:Q_quantum_numbers}.
%
%
\begin{table}[t]
  \caption{Quantum numbers of the axion multiplet $\Phi$ and 
    the heavy KSVZ quark multiplets $Q_{1,2}$ considered in this work.}
\label{Tab:Q_quantum_numbers}
\begin{center}
  \renewcommand{\arraystretch}{1.25}
\begin{tabular*}{3.25in}{@{\extracolsep\fill}rcl}
\hline
chiral multiplet           
& U(1)$_{\PQ}$ 
& (SU(3)$_\Color$,\,SU(2)$_\Weak$)$_{\Hypercharge}$
\\ \hline
$\Phi\,\,=\,\,\phi\,\,+\,\sqrt{2}\chi\theta+F_{\Phi}\theta\theta$                
& +1 
& ($\bf{1}$,\,$\bf{1}$)$_0$
\\
$Q_1=\widetilde{Q}_1+\sqrt{2}q_1\theta+F_1\theta\theta$                 
& -1/2 
& ($\bf{3}$,\,$\bf{1}$)$_{+e_Q}$
\\
$Q_2=\widetilde{Q}_2+\sqrt{2}q_2\theta+F_2\theta\theta$                 
& -1/2 
& ($\bf{3^*}$,\,$\bf{1}$)$_{-e_Q}$
\\
\hline
\end{tabular*}
\end{center}
\end{table}
%
%
With the 2-component fields listed in that table, the axino and the
heavy KSVZ quarks are described by the respective 4-component fields,
\begin{equation}
        \axino = \begin{pmatrix}\chi \\ \bar{\chi}\end{pmatrix}
        \quad \mbox{and} \quad
        Q =  \begin{pmatrix} q_1 \\ \bar{q}_2 \end{pmatrix}
\ .
\end{equation}
For the heavy KSVZ (s)quark masses, we use the SUSY limit
$M_{\widetilde{Q}_{1,2}}=M_Q=y\langle\phi\rangle=y f_a / \sqrt{2}$
with both $y$ and $\langle\phi\rangle$ taken to be real by field redefinitions.%
\footnote{To avoid ambiguities related to the $f_a$ definition,
  $f_a=\sqrt{2}\langle\phi\rangle$, the effective axion interactions
  obtained from~(\ref{Eq:SuperpotentialPQ}) after integrating out the
  heavy KSVZ fields are given in Sect.~3 of
  Ref.~\cite{Freitas:2009fb}.}
Accordingly, the phenomenological constraint~(\ref{Eq:f_a_axion})
implies, for $y=\Order(1)$, a large mass hierarchy between the KSVZ
fields and the weak and the soft SUSY mass scales,
\begin{equation}
M_{Q/\widetilde{Q}_{1,2}}\gg \mZ, m_{\SUSY} 
\ .
\label{Eq:MassHierarchy}
\end{equation}

While our study can easily be generalized, we focus on the case in
which the NLSP is a purely right-chiral lighter stau,
$\stauone=\stauR$, which is a good approximation at least for small
$\tan\beta$.
Its coupling to neutralinos is then dominated by the bino coupling.
For further simplicity, we assume that mixing in the neutralino sector
is such that one of the neutralino states is an (almost) pure bino.
In fact, our cosmological considerations rely on a spectrum in which
that state is the lightest neutralino, $\neutralino=\bino$, while our
results for the $\stauR$ decays are not restricted to this case.
We treat the axino mass $\maxino$ as a free parameter which is bounded
from above by the stau NLSP mass $\mstau$.

In the considered R-parity conserving setting, the lifetime of the
$\stauR$ NLSP is governed by the decay $\stauR\to\tau\axino$,
\begin{equation}
  \tau_{\stau} 
  =
  1/\Gamma_{\tot}^{\stauR}
  \approx
  1/\Gamma(\stauR\to\tau\axino)
  \, ,
\label{Eq:tau_stau}
\end{equation}
where the dominant contributions to the partial width
$\Gamma(\stauR\to\tau\axino)$ occur at the two-loop level.%
\footnote{As noted
  in~\cite{Covi:2004rb,Brandenburg:2005he,Schilling:2005dr,Freitas:2009fb},
  the 3-body decay $\stauR\to\tau\axino\gamma$ occurs already at the
  1-loop level, but for most of the phase space is subdominant.
  Moreover, compared to the 3-body decay, the 4-body decay modes
  $\stauR\to\tau\axino\gamma\gamma,\,\tau\axino l^+
  l^-,\,\tau\axino\quark\antiquark$ are suppressed by an additional
  factor of $\alpha$; cf.~Fig.~\ref{Fig:Stau_Four-Body} below.}
Using a heavy mass expansion in powers of
$1/f_a$~\cite{Schilling:2005dr}, we have recently calculated the
leading term that governs~(\ref{Eq:tau_stau})~\cite{Freitas:2009fb}.

\section{Energy release in stau decays}
\label{Sec:Energy_Release}

Decays of the $\stauR$ NLSP into the $\axino$ LSP are associated with
the emission of Standard Model particles.
If injected at cosmic times $t\gtrsim 100~\seconds$, i.e., during or
after BBN, those particles can reprocess the primordial light elements
significantly via hadrodissociation and/or photodissociation.
At earlier times, energetic particles are stopped efficiently through
electromagnetic interactions so that
hadrodissociation/photodissociation becomes an important issue only
for $\tau_{\stau}\gtrsim 100~\seconds$; see
Ref.~\cite{Kawasaki:2004qu,Jedamzik:2006xz,Cyburt:2009pg} and
references therein.%
\footnote{The presence of additional slow hadrons can also affect BBN
  through proton--neutron interconversion
  processes~\cite{Reno:1987qw}. The associated limits are not
  considered since they are typically milder;
  cf.~\cite{Kawasaki:2008qe} for a discussion in $\gravitino$ LSP
  scenarios with a stau NLSP.}

In this section, we explore the (average) electromagnetic/hadronic
energy emitted in a single $\stauR$ decay: $\epsilon_{\EM/\HAD}$. We
will use this quantity below to investigate whether successful BBN
predictions are preserved.

The electromagnetic energy release $\epsilon_{\EM}$ is governed by the
tau emitted in the main decay channel $\stauR\to\tau\axino$ with an
energy $E_\tau =(\mstau^2-\maxino^2+m_{\tau}^2)/(2\mstau)$ in the rest
frame of the $\stauR$. For stau decays at cosmic times $t\gtrsim
100~\seconds$, the emitted tau decays before interacting
electromagnetically.
As each $\tau$ decays into at least one $\nu$, which does not interact
electromagnetically, only a fraction of $E_\tau$
contributes~\cite{Feng:2003uy,Steffen:2006hw}. We use the conservative
estimate
\begin{equation}
  \epsilon_{\EM} 
  = 0.3\, E_\tau 
  = 0.3\, \frac{\mstau^2-\maxino^2+m_{\tau}^2}{2\mstau}
\label{Eq:epsEM}
\end{equation}
to avoid that the electromagnetic BBN constraints presented in
Sect.~\ref{Sec:ConstraintsPQScale} are overly restrictive.

The hadronic energy release $\epsilon_{\HAD}$ is governed by the
quark--antiquark pair emitted in the 4-body decay
$\stauR\to\tau\axino\quark\antiquark$.
Mesons from decays of the $\tau$'s emitted in the main decay channel
$\stauR\to\axino\tau$ typically decay before interacting
hadronically~\cite{Kawasaki:2004qu}.
In fact, since mesons typically decay before interacting with the
background nuclei, only the nucleons originating from hadronization of
the $\quark\antiquark$ pair need to be taken into account for
$\tau_{\stau}\gtrsim 100~\seconds$. We thus consider only
$\quark\antiquark$ pairs with an invariant mass $m_{\quark\antiquark}$
above the mass of a pair of nucleons, 
$m_{\quark\antiquark}>m_{\quark\antiquark}^{\mathrm{cut}}=2~\GeV$,
when calculating
\begin{align}
    \epsilon_{\HAD}  
        \equiv {1 \over \Gamma_{\tot}^{\stauR}}
        \int_{m_{\quark\antiquark}^{\mathrm{cut}}}^{\mstau-\maxino-m_{\tau}}
        \!\!\!\!\!\!\!\!
        dm_{\quark\antiquark}\,m_{\quark\antiquark}
        {d\Gamma(\stauR\to\tau\axino\quark\antiquark)
        \over dm_{\quark\antiquark}} 
        \ .
\label{Eq:epsHAD}
\end{align}
According to the value of $m_{\quark\antiquark}$, all quark flavors
that can occur in the final state are taken into account in the
differential decay rate
$d\Gamma(\stauR\to\tau\axino\quark\antiquark)/dm_{\quark\antiquark}$.

\begin{figure}[t!]
\includegraphics[width=.48\textwidth]{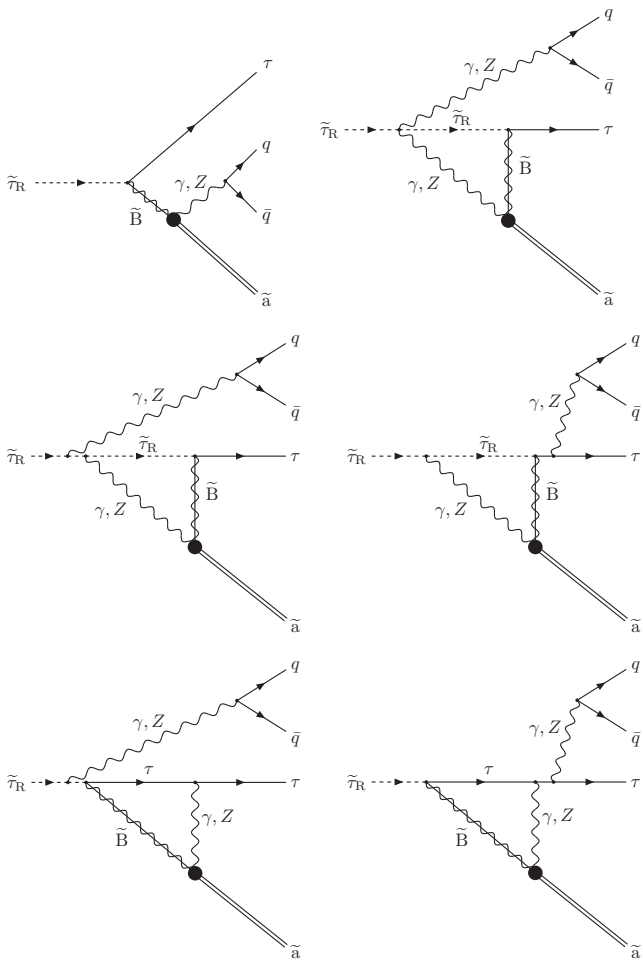}
\caption{Feynman diagrams of the dominant contributions to the stau
  NLSP decay $\stauR\to\tau\axino\quark\antiquark$ in a SUSY hadronic
  axion model. Here the lightest neutralino is assumed to be a pure
  bino $\neutralino=\bino$ and the tau mass is neglected. The blob
  represents the loops involving the heavy KSVZ (s)quark fields with
  quantum numbers given in Table~\ref{Tab:Q_quantum_numbers}; cf.\
  Fig.~2 in Ref.~\cite{Freitas:2009fb}.}
\label{Fig:Stau_Four-Body}
\end{figure}

The Feynman diagrams of the dominant contributions to the 4-body decay
$\stauR\to\tau\axino\quark\antiquark$ are illustrated in
Fig.~\ref{Fig:Stau_Four-Body}, where the blobs represent heavy KSVZ
(s)quark loops; cf.\ Fig.~2 of Ref.~\cite{Freitas:2009fb}.
%
%
As in the calculation of the 2-body decay~\cite{Freitas:2009fb}, we
work in the limit $m_{\tau}\to 0$ which is justified since $m_\tau \ll
\mstau$. 
Taking advantage of~(\ref{Eq:MassHierarchy}), we use again the method
of heavy mass expansion to expand in heavy (s)quark masses, keeping
only the leading terms $\propto 1/f_a$ in the calculation of the decay
amplitude.

Although the decay first occurs at the 1-loop level, the shown 2-loop
diagrams are also important as their contribution to the amplitude is
enhanced by large logarithms $\ln(y f_a/\sqrt{2}\mstau) \simeq 20$,
e.g., for $\mstau/y=100~\GeV$ and $f_a=10^{11}\,\GeV$. This
enhancement largely compensates for the loop suppression factor
governed by the fine-structure constant $\alpha=e^2/(4\pi)$.
Therefore, only the leading logarithmic term of the 2-loop
contribution is kept.

There are more diagrams at the 2-loop level than those shown as part
of Fig.~\ref{Fig:Stau_Four-Body}. One group has a $\gamma,Z$ line that
splits into the $\quark\antiquark$ pair after originating either from
the $\stauR$ or $\tau$ in the loop or from one of the heavy (s)quarks
in the blob. However contributions of these diagrams are canceling or
not enhanced by large logarithms and therefore neglected in our
calculation. Similarly, there are diagrams with gluons (more than one
because of color conservation) emitted by the heavy (s)quarks in the
blob at higher order in the strong coupling, turning into nucleons.
These are also highly suppressed and thus neglected.

\begin{figure}[t!]
\epsfig{figure=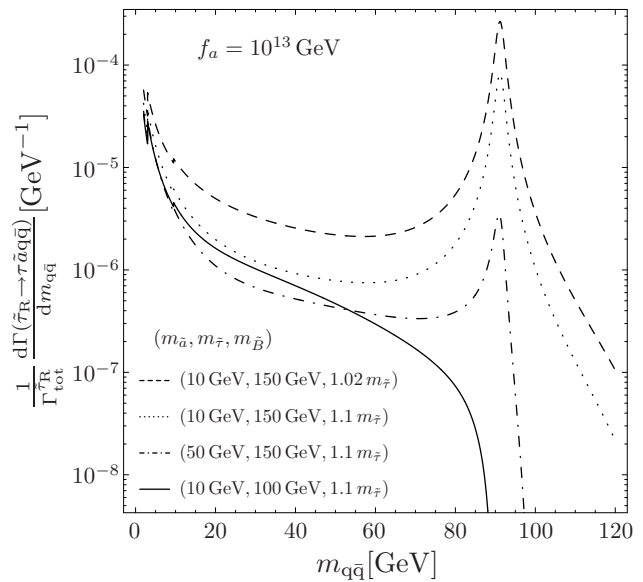, width=.46\textwidth}
\caption{Energy spectrum of the quark--antiquark pair emitted in the
  four-body decay $\stau_{\mathrm R}\to\tau\axino\quark\antiquark$
  with an invariant mass $m_{\quark\antiquark}$, normalized to the
  total decay rate
  $\Gamma_{\tot}^{\stauR}\approx\Gamma(\stauR\to\tau\axino)$.  The
  shown quantity is independent of $e_Q$. The other parameters are set
  to $f_a=10^{13}\,\GeV$ and $y=1$.}
\label{Fig:EnergySpectrum}
\end{figure}

In Fig.~\ref{Fig:EnergySpectrum}, our results for the differential
decay rate
$d\Gamma(\stauR\to\tau\axino\quark\antiquark)/dm_{\quark\antiquark}$
normalized to the total decay rate $\Gamma_{\tot}^{\stauR}$ are shown
for $(\maxino,\,\mstau,\,\mbino)$ combinations of
$(10~\GeV,\,100~\GeV,\,1.1\,\mstau)$, $(50~\GeV,$
$150~\GeV,\,1.1\,\mstau)$, $(10~\GeV,\,150~\GeV,\,1.1\,\mstau)$, and
$(10~\GeV,\,150~\GeV,\,1.02\,\mstau)$ by the solid, dash-dotted,
dotted, and dashed curves, respectively.
%
%
The shown quantity is independent of $e_Q$. The other parameters are
set to $f_a=10^{13}\,\GeV$ and $y=1$.  The kinks at
$m_{\quark\antiquark}\simeq 3~\GeV$ and $10~\GeV$ mark the respective
thresholds for the production of $\cquark\anticquark$ and
$\bquark\antibquark$ pairs.
Moreover, one can see the contribution of the Z-boson resonance at
$m_{\quark\antiquark}=\mZ$, which is present for
$\Delta m\equiv\mstau-\maxino-m_{\tau}>\mZ$.
As part of the integrand in~(\ref{Eq:epsHAD}), the shown energy
spectrum of the emitted $\quark\antiquark$ pair is the crucial
quantity in the calculation of $\epsilon_{\HAD}$.
Based on the numerical results from the integration
in~(\ref{Eq:epsHAD}), we obtain the hadronic BBN constraints presented
in Sects.~\ref{Sec:HadBBNaxino} and~\ref{Sec:ConstraintsPQScale}.
We note that a more precise calculation of the constraints would
require, in addition to 
$d\Gamma(\stauR\to\tau\axino\quark\antiquark)/dm_{\quark\antiquark}$,
a treatment of the fragmentation of the quarks into hadrons and of the
propagation of the resulting hadron spectra when computing the
abundances of primordial light elements; cf.~\cite{Kawasaki:2004qu,
  Kohri:2005wn, Bailly:2008yy, Cyburt:2009pg}.

\section{Cosmological setting}
\label{Sec:Cosmological_Setting}

In addition to $\tau_{\stau}$ and $\epsilon_{\EM/\HAD}$, the stau NLSP
yield prior to decay,
$Y_{\stau}\equiv n_{\stauR}/s$,
is another quantity that is crucial for our study of cosmological
constraints. Here $s$ denotes the entropy density and
$n_{\stauR}=n_{\stauR^+}+n_{\stauR^-}=2\,n_{\stauR^-}$ the total
$\stauR$ number density prior to decay for an equal number of
positively and negatively charged $\stauR$s.
In contrast to $\tau_{\stau}$ and $\epsilon_{\EM/\HAD}$, $Y_{\stau}$
depends on the thermal history of the early Universe.
In this Letter we assume a standard thermal history with a reheating
temperature after inflation in the range:
$T_{\decoupling}^{\stauR}<\TR< f_a$,
where $T_{\decoupling}^{\stauR}\sim\mstau/25$ is the temperature at
which the $\stauR$ NLSP species decouples from the primordial plasma.
This has a number of implications: 
\begin{itemize}
\item[(i)] Focussing on a standard thermal history, we assume that
  effects of the saxion---the bosonic partner of the axino that
  appears in addition to the axion---are negligible. For a
  non-standard thermal history associated with significant late
  entropy production in saxion
  decays~\cite{Kim:1992eu,Lyth:1993zw,Chang:1996ih,Hashimoto:1998ua},
  cosmological constraints~\cite{Kawasaki:2007mk} including those
  considered in this work can be affected depending on the saxion
  properties.
\item[(ii)] For $\TR<f_a$, no PQ symmetry restoration takes place
  after inflation. In fact, we assume that the PQ symmetry was broken
  before inflation and not restored afterwards. For large $f_a$ such
  that axions are never in thermal equilibrium with the primordial
  plasma, the relic axion density $\Omega_a$ is then governed by the
  initial misalignment angle $\Theta_i$ of the axion field with
  respect to the CP-conserving position;
  cf.~\cite{Sikivie:2006ni,Raffelt:2006cw} and references therein.
  This allows us to keep the presented constraints conservative by
  assuming $\Omega_a\ll\OmegaDM$ which is possible even for $f_a$
  above $10^{14}\,\GeV$ since $\Theta_i$ can be sufficiently small.
\item[(iii)] For $T_{\decoupling}^{\stauR}<\TR$, the $\stauR$ NLSP
  decouples as a weakly interacting massive particle before its decay
  into the axino LSP so that $Y_{\stau}$ is the thermal relic stau
  abundance, which does not depend on $\TR$.
\end{itemize}

The thermal relic stau abundance prior to decay can be calculated
numerically. Its value depends on details of the SUSY model such as
the mass splitting among the lightest Standard Model
superpartners~\cite{Asaka:2000zh} or the left-right mixing of the stau
NLSP~\cite{Ratz:2008qh,Pradler:2008qc}.
With the focus on the $\stauR$ NLSP setting in this Letter, we
consider three characteristic approximations:
\begin{align}
  Y_{\stau}
  \simeq
  \kappa \times 10^{-12}
  \left(\frac{m_{\stau}}{1~\TeV}\right),
  \quad
  \kappa = 0.7,\,1.4,\,2.8, 
\label{Eq:Ystau}
\end{align}
where $\kappa$ accounts for typical differences in the annihilation
processes of the sleptons. 
The value $\kappa=0.7$ corresponds to the case with $\mbino=1.1\,\mstau$
and $\mstau\ll m_{\sel,\smu}$, in which primordial stau annihilation
involves only staus in the initial
state~\cite{Asaka:2000zh}.%
\footnote{The bino mass $\mbino=1.1\,\mstau$ considered in
  Ref.~\cite{Asaka:2000zh} represents a typical mass splitting in
  regions with $\mbino>\mstau$ encountered in scenarios such as the
  constrained MSSM (CMSSM).}
 The yield associated with $\kappa=1.4$ is encountered if there is either
additional stau--slepton coannihilation corresponding to
$\mstau\lesssim m_{\sel,\smu}<1.1\,\mstau$~\cite{Asaka:2000zh} or
additional stau--bino coannihilation corresponding to
$\mstau\lesssim\mbino<1.1\,\mstau$ (cf.\ $Y_{\stau}$ contours close to
the dashed line in the right panel of Fig.~3 in
Ref.~\cite{Pradler:2006hh}).
For an approximate degeneracy of $\mstau$ with both $m_{\sel,\smu}$
and $\mbino$, simultaneous stau--slepton--bino coannihilation can lead
to an even larger $\kappa=2.8$ in~(\ref{Eq:Ystau}) (cf.\ $Y_{\stau}$
contours close to the dashed line in the left panel of Fig.~3 of
Ref.~\cite{Pradler:2006hh}).

A non-thermally produced (NTP) axino density~\cite{Bonometto:1993fx,Covi:1999ty,Covi:2001nw,Covi:2004rb}
\begin{equation}
        \Omega_{\axino}^{\NTP} h^2
        = 
        m_{\axino}\, Y_{\stau}\, s(T_0) h^2 / \rho_{\mathrm{c}}
\label{Eq:AxinoDensityNTP}
\end{equation}
emerges since one axino LSP is emitted in each stau NLSP decay;
$\rho_c/[s(T_0)h^2]=3.6\times10^{-9}\,\GeV$~\cite{Amsler:2008zz}.
This density must not exceed the dark matter
density~\cite{Spergel:2006hy}
\begin{equation}
  \OmegaDM^{3\sigma}h^2=0.105^{+0.021}_{-0.030}
  \, ,
\label{Eq:OmegaDM}
\end{equation}
where a nominal $3\sigma$ range is indicated and where
$h=0.73^{+0.04}_{-0.03}$ denotes the Hubble constant in units of
$100~\km\,\Mpc^{-1}\seconds^{-1}$.
Additionally taking into account the thermally produced axino density
$\Omega_{\axino}^{\TP}$~\cite{Brandenburg:2004du,Freitas:2009fb} and
the axion density $\Omega_a$, one obtains the dark matter constraint
on the stau abundance prior to decay: $Y_{\stau}\leq
Y_{\stau\,\CDM}^{\max}$ with
\begin{equation}
  Y_{\stau\,\CDM}^{\max}
  \!=\! 
  4.5\times 10^{-11}\!
  \left(\frac{\OmegaDM\!-\!\Omega_{\axino}^{\TP}\!\!-\!\Omega_a}{0.126/h^2}\!\right)\!
  \left(\frac{10~\GeV}{\maxino}\right)\!
  .
\label{Eq:DMconstraint}
\end{equation}
For the most conservative case
$\Omega_{\axino}^{\TP}+\Omega_a\ll\OmegaDM$,
we have illustrated the constraint on $\maxino$ and $\mstau$ obtained
by confronting $Y_{\stau}$ with~(\ref{Eq:DMconstraint}) already in
Ref.~\cite{Freitas:2009fb}.  

For $\tau_{\stau}>10^3\,\seconds$, additional upper limits on
$Y_{\stauR^-}=Y_{\stau}/2$ occur since the negatively charged
$\stauR^-$s can form $(\Hefour\,\stauR^-)$ and $(\beetm\,\stauR^-)$
bound states and can thereby catalyze primordial $^6$Li and $^9$Be
production in excess of observationally inferred
limits~\cite{Pospelov:2006sc,Pospelov:2007js,Pospelov:2008ta}.%
\footnote{We note that Ref.~\cite{Kamimura:2008fx} has questioned the
  efficiency of the catalyzed production of $\ben$ obtained in
  Refs.~\cite{Pospelov:2007js,Pospelov:2008ta}, and announced a
  further clarification of this point in the future.}
In fact, the CBBN constraints on $\maxino$ and $\mstau$ obtained by
confronting $Y_{\stau}$ with the CBBN-induced $\tau_{\stau}$-dependent
upper limits given in Ref.~\cite{Pospelov:2008ta} have been presented
for the first time in Ref.~\cite{Freitas:2009fb}.

\section{Hadronic BBN Constraints}
\label{Sec:HadBBNaxino}

%
\begin{figure*}[t!]
\includegraphics[width=.46\textwidth]{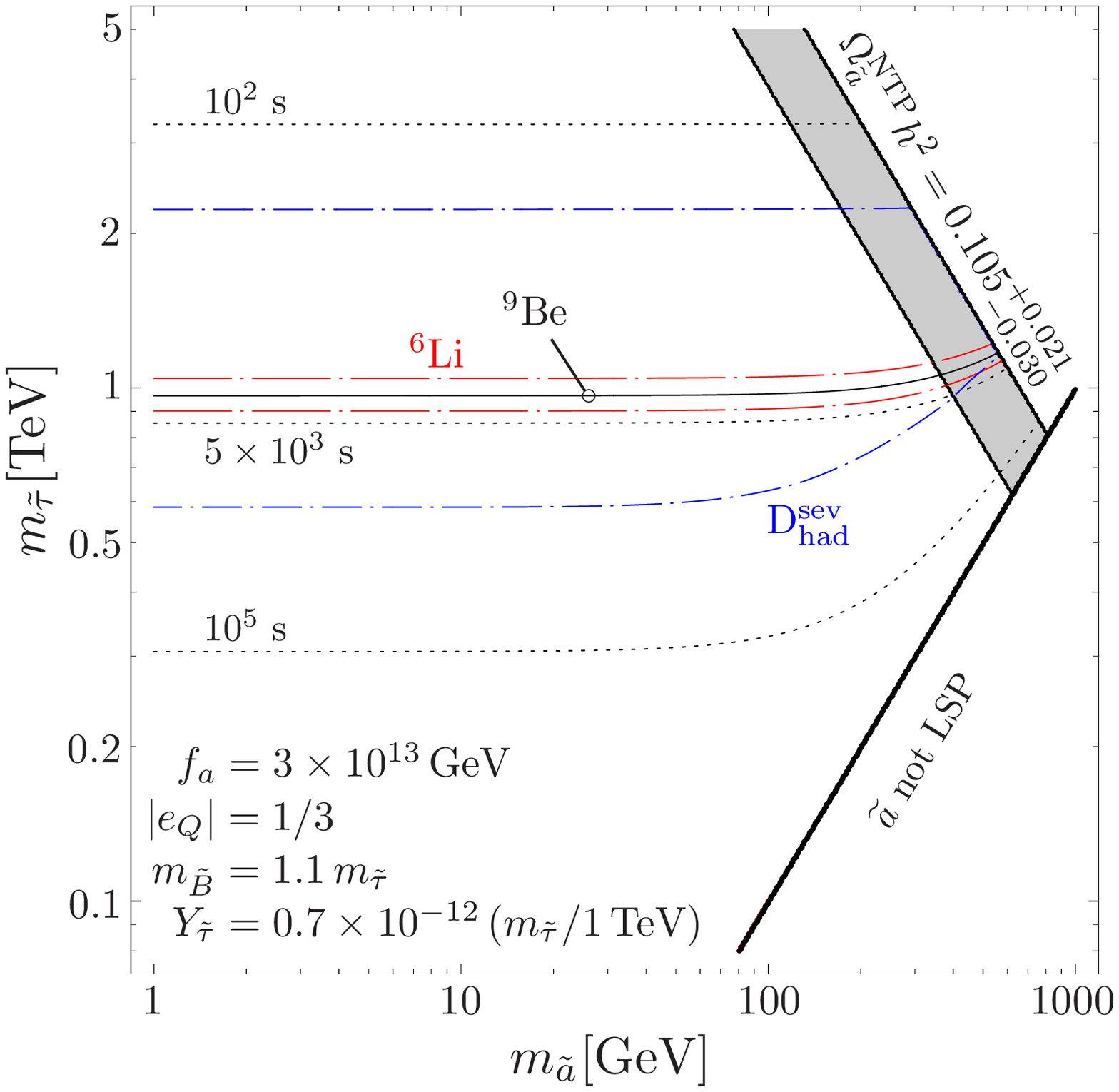}
\hfill
\includegraphics[width=.46\textwidth]{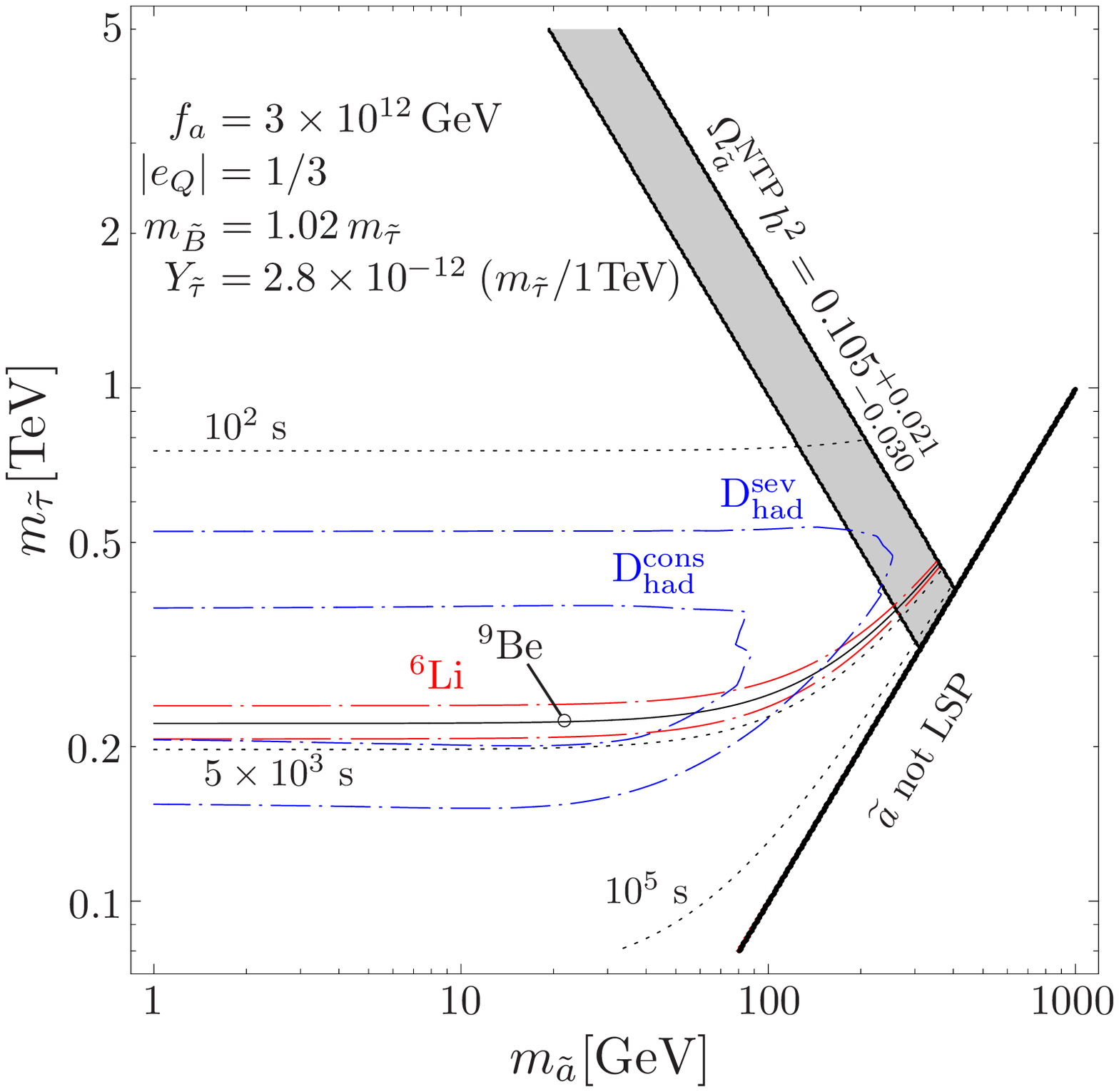}
\vskip -0.5cm
\makebox[.46\textwidth][l]{\textbf{(a)}}\hfill
\makebox[.46\textwidth][l]{\textbf{(b)}}\hfill
\caption{Cosmological constraints on the masses of the $\axino$ LSP
  and the $\stauR$ NLSP for (a)~$f_a=3\times 10^{13}\,\GeV$,
  $\mbino=1.1\,\mstau$, $Y_{\stau}$ given by~(\ref{Eq:Ystau}) with
  $\kappa=0.7$ and (b)~$f_a=3\times 10^{12}\,\GeV$,
  $\mbino=1.02\,\mstau$, $Y_{\stau}$ given by~(\ref{Eq:Ystau}) with
  $\kappa=2.8$. In both panels, $|e_Q|=1/3$ and $y=1$. The hadronic BBN
  constraints associated with~(\ref{Eq:D_sev}) and~(\ref{Eq:D_cons})
  disfavor the regions enclosed by the respective short-dash-dotted
  (blue) lines.  The CBBN constraints associated with~(\ref{Eq:LiSix})
  and~(\ref{Eq:BeNine}) disfavor the regions below the
  long-dash-dotted (red) and the solid lines. Contours of
  $\tau_{\stau}=10^2$, $5\times 10^3$, and $10^5\,\seconds$ are shown
  by the dotted lines. On (above) the gray band,
  $\Omega_{\axino}^{\NTP}\in\Omega_{\CDM}^{3\sigma}$
  ($\Omega_{\axino}^{\NTP}h^2>0.126$). The region with
  $\maxino>\mstau$ is not considered as we focus on the $\axino$ LSP
  case.}
\label{Fig:HAD_BBN1}
\end{figure*}

Hadronic energy injection can affect the abundance of primordial
deuterium substantially via hadrodissociation of helium-4.
In fact, we focus on the constraint on hadronic energy release imposed
by the primordial abundance of $\dm$ in this section.
While additional constraints on late energy injection are imposed by
the primordial abundances of $^4$He, $^3$He/D, $^7$Li, and
$^6$Li/$^7$Li~\cite{Sigl:1995kk,Jedamzik:1999di,Jedamzik:2004er,Kawasaki:2004qu,Jedamzik:2006xz,Cyburt:2006uv,Cyburt:2009pg},
the hadronic $\dm$ constraint is the dominant one in the region
allowed by the CBBN constraints, $\tau_{\st}\lesssim 10^3~\seconds$.
This can be seen, e.g., in Figs.~38--41 of Ref.~\cite{Kawasaki:2004qu}
and in Figs.~6--8 of Ref.~\cite{Jedamzik:2006xz}.
Also the $\dm$ constraint on electromagnetic energy injection is not
considered in this section since it becomes relevant only for
$\tau_{\st}>10^4~\seconds$ which is already disfavored by CBBN; see
Sect.~\ref{Sec:ConstraintsPQScale} for details.

To derive the hadronic BBN constraint imposed by $\dm$, we use the
$\tau_{\NLSP}$-dependent upper limits (95\%~CL) on
\begin{align}
  \xi_{\HAD} \equiv \epsilon_{\HAD}\, Y_{\NLSP}
\label{Eq:EnergyRelease}
\end{align}
given in Fig.~9 of Ref.~\cite{Steffen:2006hw} as obtained in
Ref.~\cite{Kawasaki:2004qu} for observationally inferred primordial D
abundances of
\begin{eqnarray}
  \!\!\!\!\!\!\!\!\!\!\!\!
  & \dm/\Hyd|_{\mathrm{mean}} 
  &=\, 
  (2.78^{+0.44}_{-0.38})\times 10^{-5} 
  \,\,\,\, \mathrm{(severe)}
  \ ,
\label{Eq:D_sev}\\
  \!\!\!\!\!\!\!\!\!\!\!\!
  &\dm/\Hyd|_{\mathrm{high}} 
  &=\, 
  (3.98^{+0.59}_{-0.67})\times 10^{-5}
  \,\,\,\,
  \mathrm{(conservative)}
  \ . 
\label{Eq:D_cons}
\end{eqnarray}
Here $\tau_{\NLSP}=\tau_{\stau}$, $Y_{\NLSP}=Y_{\stau}$, and
$\epsilon_{\HAD}$ is given in~(\ref{Eq:epsHAD}).
With the calculated $\epsilon_{\HAD}$, upper limits on $Y_{\stau}$ can
be derived from the considered upper limits on
$\xi_{\HAD}$~\cite{Freitas:2009xxx}:
$Y_{\stau\,\HAD}^{\max}=\xi_{\HAD}^{\max}/\epsilon_{\HAD}$.
By confronting $Y_{\stau}$ with $Y_{\stau\,\HAD}^{\max}$, we then
obtain the regions in the parameter space that are disfavored by the
hadronic BBN constraints.

In Fig.~\ref{Fig:HAD_BBN1} the obtained hadronic BBN constraints are
shown by short-dash-dotted (blue) lines. The labels
$\dm_{\HAD}^{\mathrm{sev}}$ and $\dm_{\HAD}^{\mathrm{cons}}$ indicate
the constraints associated with~(\ref{Eq:D_sev})
and~(\ref{Eq:D_cons}), respectively. The regions enclosed by the
corresponding lines are disfavored by an excess of $\dm$ above the
respective observationally inferred abundance.
The absence of a $\dm_{\HAD}^{\mathrm{cons}}$ line in panel~(a) and
the difference between the $\dm_{\HAD}^{\mathrm{sev}}$ and
$\dm_{\HAD}^{\mathrm{cons}}$ lines in panel~(b) indicate the
sensitivity on the observationally inferred $\dm$ abundance.

We also show the CBBN constraints associated with the observationally
inferred limits on the respective primordial fractions of
$\Lisix$~\cite{Cyburt:2002uv,Asplund:2005yt,Jedamzik:2007qk} and
$\ben$~\cite{Pospelov:2008ta},
\bea
  \Lisix/\mathrm{H}|_{\mathrm{obs}} 
  & \leq & 
  10^{-11}\!-\!10^{-10}
  \ ,
\label{Eq:LiSix}\\
  \ben/\mathrm{H}|_{\mathrm{obs}} 
  & \leq &
  2.1\times 10^{-13}
  \ ,
\label{Eq:BeNine}
\eea
as obtained by confronting $Y_{\stauR^-}$ with the limits in Fig.~5 of
Ref.~\cite{Pospelov:2008ta}.
The range~(\ref{Eq:LiSix}) is indicated by pairs of long-dash-dotted
($\Lisix$, red) lines and~(\ref{Eq:BeNine}) by solid ($\ben$) lines.
The regions below those lines are disfavored by an excess of $\Lisix$
and $\ben$ above the respective limits.
The region with
$\Omega_{\axino}^{\NTP}\in\OmegaDM^{3\sigma}$
is indicated by the gray band, where the region above that band is the
one excluded by~(\ref{Eq:DMconstraint}) in the conservative case with
$\Omega_{\axino}^{\TP}+\Omega_a\ll\OmegaDM$.
The dotted lines are contours of $\tau_{\stau}=10^2$, $5\times 10^3$,
and $10^5\,\seconds$.

\begin{figure*}[t!]
\includegraphics*[width=.46\textwidth]{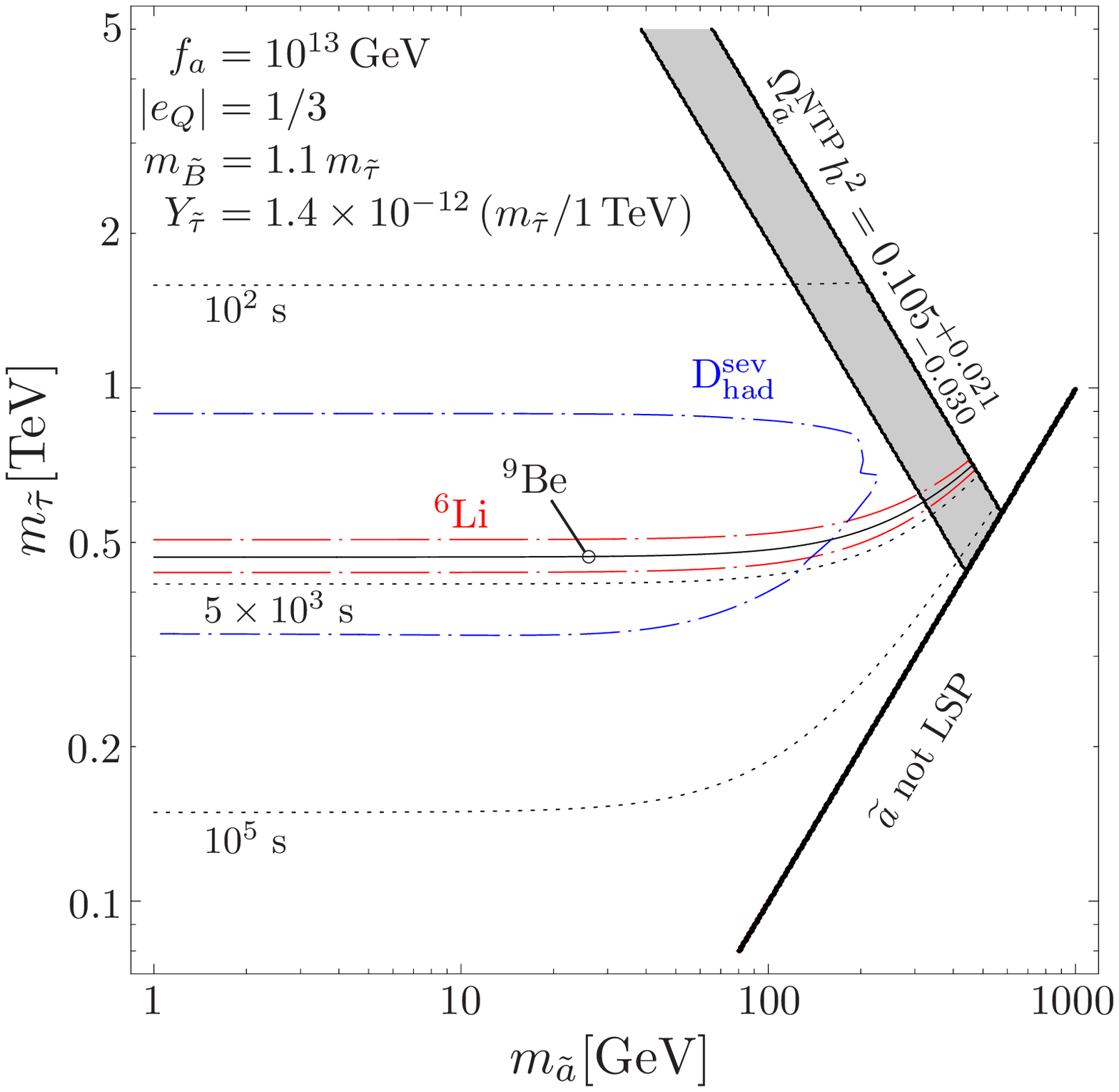}
\hfill
\includegraphics*[width=.46\textwidth]{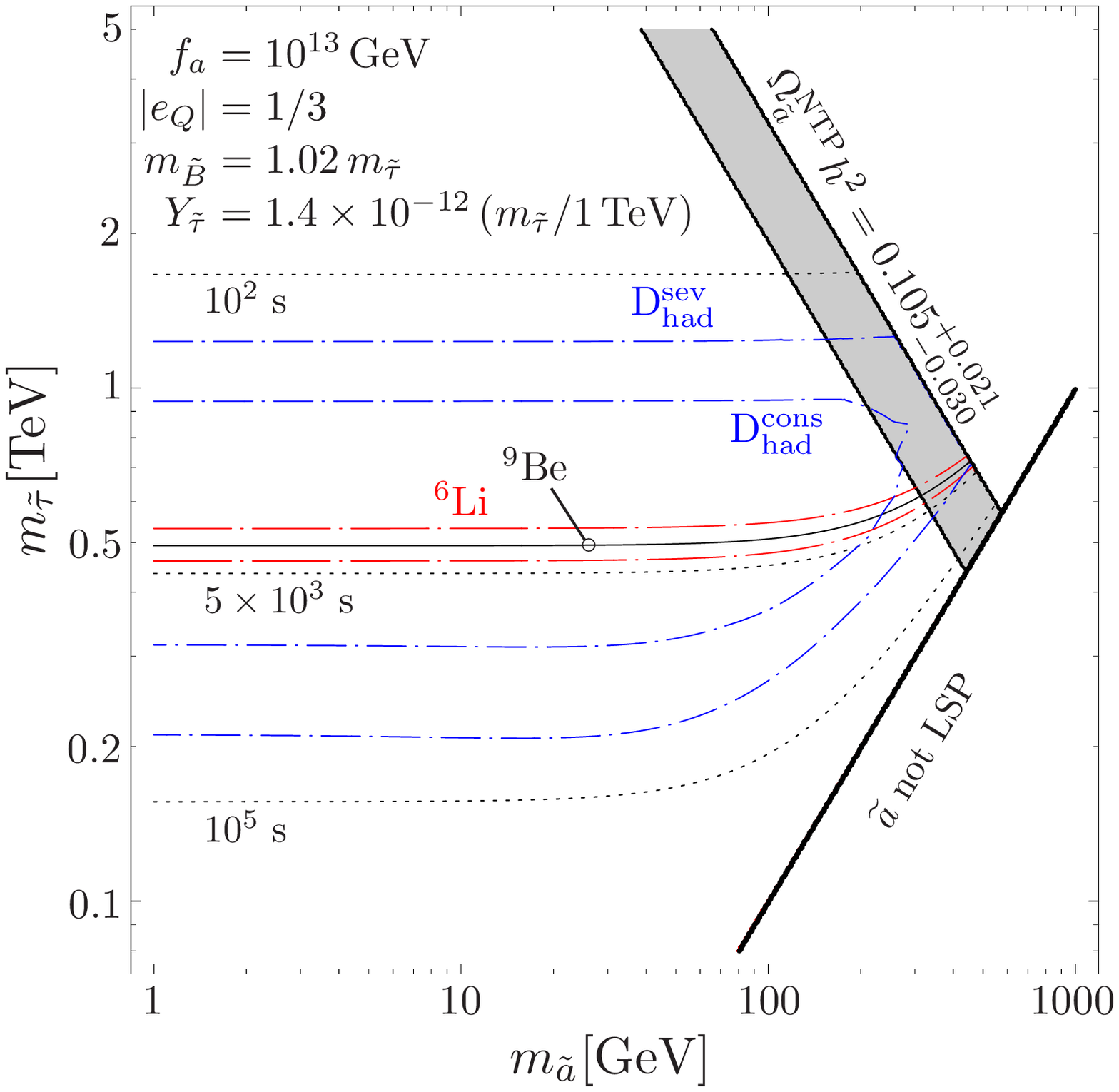}
\vskip -0.5cm
\makebox[.46\textwidth][l]{\textbf{(a)}}\hfill
\makebox[.46\textwidth][l]{\textbf{(b)}}\hfill
\caption{Cosmological constraints on the masses of the $\axino$ LSP
  and the $\stauR$ NLSP for (a)~$\mbino=1.1\,\mstau$ and
  (b)~$\mbino=1.02\,\mstau$. In both panels, $f_a=10^{13}\,\GeV$,
  $|e_Q|=1/3$, $y=1$, and $Y_{\stau}$ is given by~(\ref{Eq:Ystau})
  with $\kappa=1.4$. The BBN constraints, the dark matter constraint,
  the $\tau_{\stau}$ contours, and the other regions are indicated as
  in Fig.~\ref{Fig:HAD_BBN1}.}
\label{Fig:HAD_BBN2}
\end{figure*}

For $|e_Q|=1/3$ and $y=1$, we consider $f_a=3\times 10^{13}\,\GeV$,
$\mbino=1.1\,\mstau$, $Y_{\stau}$ given by~(\ref{Eq:Ystau}) with
$\kappa=0.7$ in panel~(a) and $f_a=3\times 10^{12}\,\GeV$, 
$\mbino=1.02\,\mstau$, $Y_{\stau}$ given by~(\ref{Eq:Ystau}) with
$\kappa=2.8$ in panel~(b).
In both panels, one finds $\mstau$ values disfavored by the hadronic
BBN constraints that are significantly larger than the ones disfavored
by the CBBN constraints.
However, for $|e_Q|=1/3$, $y=1$, $\mbino=1.1\,\mstau$ and $\kappa=0.7$,
even the more restrictive `severe' hadronic BBN constraint disappears
for $f_a\lesssim 10^{13}\,\GeV$. In contrast, the CBBN constraints
remain until $f_a\lesssim 10^{12}\,\GeV$---cf.\
Fig.~\ref{Fig:fa_BBN}(a)---given the limit $\mstau\gtrsim
80~\GeV$~\cite{Amsler:2008zz} from searches for long-lived staus at
the Large Electron Positron (LEP) collider.

For $\mbino=1.02\,\mstau$ and $Y_{\stau}$ enhanced by
stau--slepton--bino coannihilation such that $\kappa=2.8$,
Fig.~\ref{Fig:HAD_BBN1}(b) shows that `severe' and `conservative'
hadronic BBN constraints are still encountered at $f_a=3\times
10^{12}\,\GeV$. This demonstrates the sensitivity of those constraints
on the stau NLSP yield. In fact, the restrictive
$\dm_{\HAD}^{\mathrm{sev}}$ constraint remains until $f_a \lesssim
10^{12}\,\GeV$ for $\kappa=2.8$; cf.\ Fig.~\ref{Fig:fa_BBN}(b).

In the latter case, more restrictive hadronic BBN constraints are
encountered not only due to the enhanced yield but also due to the
smaller bino mass which is associated with larger values of
$\epsilon_{\HAD}$ and thereby with more restrictive
$Y_{\stau\,\HAD}^{\max}$ values. 
Fig.~\ref{Fig:HAD_BBN2} shows this more clearly for
$f_a=10^{13}\,\GeV$, $|e_Q|=1/3$, $y=1$, and $\kappa=1.4$, where
constraints and $\tau_{\stau}$ values are indicated as in
Fig.~\ref{Fig:HAD_BBN1}.
The two panels seen here only differ in the bino--stau mass ratio.
The one on the left illustrates the case where we have
$\mbino=1.1\,\mstau$ and stau--slepton coannihilation while the panel
on the right illustrates the case where we have $\mbino=1.02\,\mstau$
and stau--bino coannihilation, where the different origins of 
coannihilation only serve to give the same $Y_\stau$ in the two cases.
However, this change in bino--stau mass ratio alters the result 
sufficiently to produce much more restrictive D constraints in the 
latter case. 
We can trace this effect back to $\epsilon_\HAD$ through
Fig.~\ref{Fig:EnergySpectrum}, which illustrates the considerably
higher curve for the case of $\mbino=1.02\,\mstau$ as compared to
$\mbino=1.1\,\mstau$, with the same $\maxino$, $\mstau$, and $f_a$
values (the top two curves in the plot).
A thorough study of this and of other aspects of the hadronic BBN
constraints will be presented in a forthcoming
publication~\cite{Freitas:2009xxx}.
Our objective in the present Letter is (i)~a first presentation of the
hadronic BBN constraints in the case with the axino LSP and a charged
slepton NLSP and (ii)~to demonstrate explicitly that they can be more
restrictive than the associated CBBN constraints.

Before proceeding let us comment on the potential interplay between
late energy injection and CBBN. The CBBN limits adopted from
Ref.~\cite{Pospelov:2008ta} have been derived for an abundance of
$\dm$ obtained with standard BBN. For an increased $\dm$ abundance
from hadrodissociation of $\hef$, CBBN of $\Lisix$ and $\ben$ becomes
more efficient.
This is evident for $\Lisix$ since its catalysis proceeds via
$(\Hefour\,\stauR^-)+\deuterium\to\Lisix+\stauR^-$~\cite{Pospelov:2006sc}
and since the primordial abundance of $\dm$ stays significantly below
the one of $\Hefour$ at the relevant times (even for a maximum of
observationally tolerable hadrodissociation of $\Hefour$); cf.\
Fig.~2.4 in Ref.~\cite{Pradler:2009dr}.
An increased output of $\ben$ results from the final step of its catalysis,
$(\beetm\,\stauR^-)+n\to\ben+\stauR^-$~\cite{Pospelov:2007js,Pospelov:2008ta},
which becomes more efficient since an enhanced abundance of $\dm$
increases the number of neutrons $n$ at the relevant
times~\cite{Mukhanov:2003xs}; cf.\ Fig.~4 in
Ref.~\cite{Pospelov:2008ta}.
Moreover, the debris of hadrodissociated $\Hefour$ can hit ambient
$\Hefour$ and thereby fuse additional
$\Lisix$~\cite{Jedamzik:2004er,Kawasaki:2004qu,Jedamzik:2006xz}.
The interplay of late energy injection and CBBN will thus lead to
constraints that can only be stronger than the ones presented in this
Letter.
Aiming at conservative limits, this allows us to neglect those
intricacies which will have to be faced in future refinements of the
presented constraints.

\section{BBN Constraints on the PQ Scale}
\label{Sec:ConstraintsPQScale}

\begin{figure*}[t!]
\includegraphics*[width=.46\textwidth]{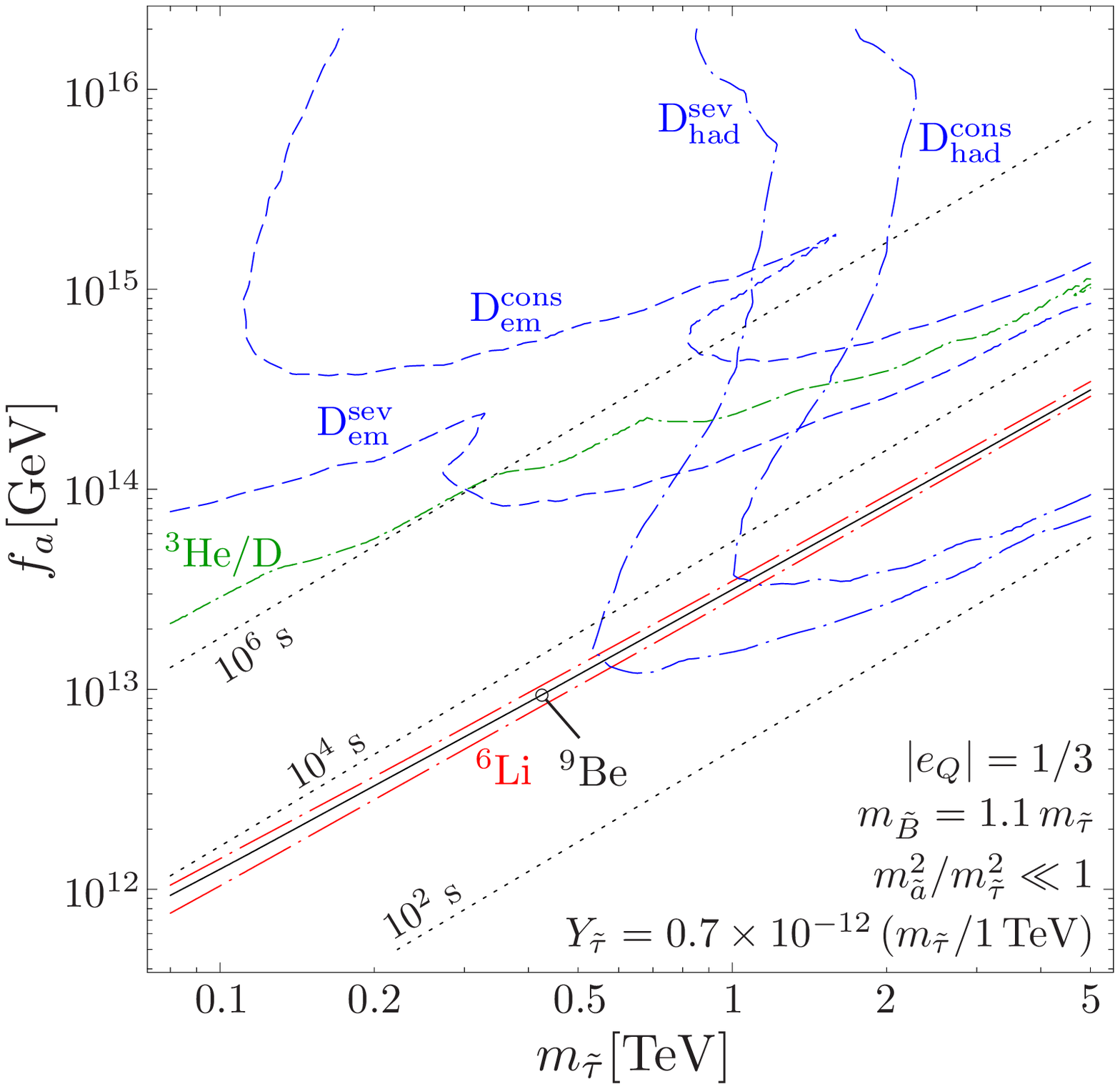}
\hfill
\includegraphics*[width=.46\textwidth]{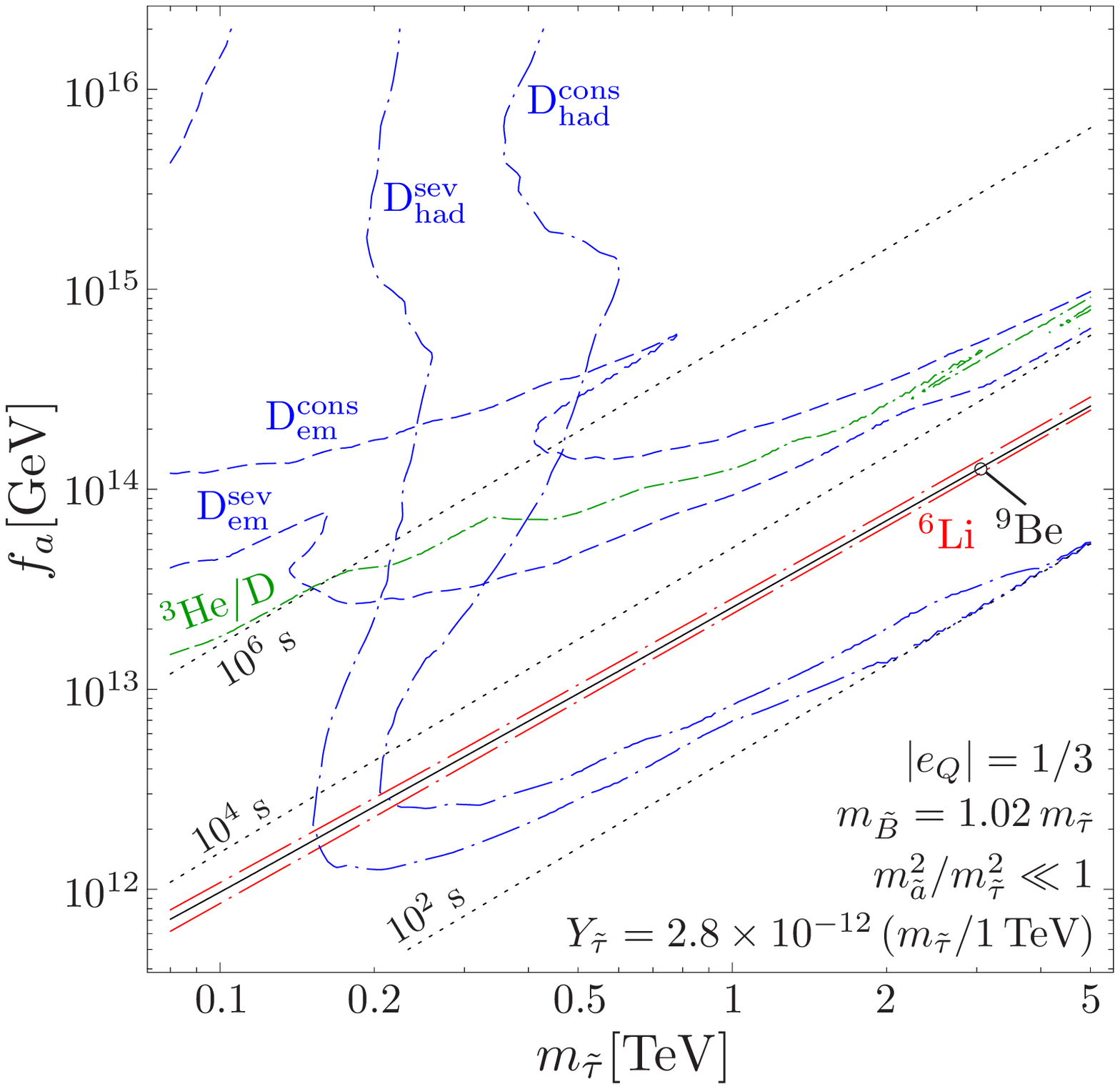}
\vskip -0.5cm
\makebox[.46\textwidth][l]{\textbf{(a)}}\hfill
\makebox[.46\textwidth][l]{\textbf{(b)}}\hfill
\caption{BBN constraints on the PQ scale $f_a$, shown for
  (a)~$\mbino=1.1\,\mstau$, $Y_{\stau}$ given by~(\ref{Eq:Ystau}) with
  $\kappa=0.7$ and (b)~$\mbino=1.02\,\mstau$, $Y_{\stau}$ given
  by~(\ref{Eq:Ystau}) with $\kappa=2.8$. In both panels,
  $\maxino^2/\mstau^2\ll 1$, $|e_Q|=1/3$, and $y=1$. The hadronic BBN
  constraints associated with~(\ref{Eq:D_sev}) and~(\ref{Eq:D_cons})
  disfavor the regions in the upper right-hand corner enclosed by the
  respective short-dash-dotted (blue) lines. Electromagnetic BBN
  constraints associated with $\dm$ disfavor the upper regions
  enclosed by the respective dashed (blue) lines and the ones
  associated with $\het/\dm$ the region above the double-dash-dotted
  (green) line. The regions above the long-dash-dotted (red) and the
  solid lines are disfavored by the CBBN constraints associated
  with~(\ref{Eq:LiSix}) and~(\ref{Eq:BeNine}). Contours of
  $\tau_{\stau}=10^2$, $10^4$, and $10^6\,\seconds$ are shown by the
  dotted lines.}
\label{Fig:fa_BBN}
\end{figure*}

In this section we show that the BBN constraints associated with
hadronic and electromagnetic energy release impose new upper limits on
the PQ scale $f_a$. In agreement with the results of the previous
section, we find that those limits can be substantially more
restrictive than the ones imposed by the CBBN
constraints~\cite{Freitas:2009fb}.

Fig.~\ref{Fig:fa_BBN} presents the new $f_a$ limits together with
the CBBN limits for $\maxino^2/\mstau^2\ll 1$, $|e_Q|=1/3$, and $y=1$.
In panel~(a) we have the generic case of $\mbino=1.1\,\mstau$ and
$Y_{\stau}$ given by~(\ref{Eq:Ystau}) with $\kappa=0.7$ while in
panel~(b) we have $\mbino=1.02\,\mstau$ and $\kappa=2.8$ representing
the case with simultaneous stau--slepton--bino coannihilation.
Contours of $\tau_{\stau}=10^2$, $10^4$, and $10^6\,\seconds$ are
shown by the dotted lines.
Above the long-dash-dotted (red) and the solid lines, CBBN of $\Lisix$
and $\ben$ is in excess of the respective limits~(\ref{Eq:LiSix})
and~(\ref{Eq:BeNine}).
The new hadronic BBN constraints associated with~(\ref{Eq:D_sev})
and~(\ref{Eq:D_cons}) disfavor the regions in the upper right-hand
corner enclosed by the respective short-dash-dotted (blue) lines.
Note that those constraints are provided only for
$\tau_{\stau}\ge 100~\seconds$
since we have not considered the typically milder limits associated
with proton--neutron interconversion processes~\cite{Reno:1987qw}
which become relevant for smaller
$\tau_{\stau}$~\cite{Kawasaki:2004qu,Jedamzik:2006xz,Cyburt:2009pg}.
Nevertheless, the hadronic BBN constraints place limits on the PQ
scale $f_a$ that become clearly more restrictive than the CBBN-induced
limits towards large $m_{\stau}$ and/or large $Y_{\stau}$.
In fact, the hadronic BBN constraint on $f_a$ can be the dominant one
already in a mass range, $m_{\stau}<1~\TeV$, that is promising for a
discovery of a long-lived stau at the LHC.

While the above sets of BBN constraints correspond to the ones shown
in the previous section (cf.\ Fig.~\ref{Fig:HAD_BBN1}), we also
indicate in Fig.~\ref{Fig:fa_BBN} the electromagnetic BBN constraints
imposed by primordial $\dm$ and $\het/\dm$.
Our derivation of the electromagnetic BBN constraints proceeds as
outlined for the hadronic ones in Sect.~\ref{Sec:HadBBNaxino} but
relies on the conservative $\epsilon_{\EM}$~(\ref{Eq:epsEM}) and on
upper limits on
$\xi_{\EM}\equiv\epsilon_{\EM}\,Y_{\stau}$.
Accordingly, we obtain the shown $\dm_{\EM}^{\mathrm{sev}}$ and
$\het/\dm$ constraints from the respective limits given in Fig.~42 of
Ref.~\cite{Kawasaki:2004qu} and the $\dm_{\EM}^{\mathrm{cons}}$
constraint from the respective limit given in Fig.~6 of
Ref.~\cite{Cyburt:2002uv}.
Those $\dm_{\EM}^{\mathrm{sev}/\mathrm{cons}}$ and $\het/\dm$
constraints disfavor the upper regions enclosed by the respective
dashed (blue) lines and the regions above the double-dash-dotted
(green) lines in Fig.~\ref{Fig:fa_BBN}.

Fig.~\ref{Fig:fa_BBN} shows that the electromagnetic BBN constraints
appear only for $\tau_{\st}>10^4~\seconds$ thereby excluding regions
already disfavored by CBBN.
Nevertheless, they support the finding that, e.g., values of the PQ
scale at the scale of grand unification, $f_a\sim 10^{16}\,\GeV$, will
be in conflict with successful BBN in the considered scenarios once a
long-lived charged slepton is observed at the LHC.

Before closing let us discuss the robustness of the shown $f_a$ limits
and address important sensitivities:
\begin{itemize}
\item By considering $\maxino^2/\mstau^2\ll 1$, the CBBN-imposed $f_a$
  limits are conservative limits. Those constraints become more
  restrictive for $\maxino\to\mstau$. This is different for
  constraints associated with late energy injection, where any bound
  can be evaded for a finely tuned $\maxino$--$\mstau$ degeneracy
  leading to $\epsilon_{\HAD/\EM}\to 0$.
\item The $f_a$ limits are sensitive to $Y_{\stau}$.
  In settings with a sizable left-right stau mixing, an exceptionally
  small $Y_{\stau}$ is possible such that even the CBBN constraints
  may be respected~\cite{Ratz:2008qh,Pradler:2008qc}.
\item The $f_a$ limits depend on the quantum numbers of the heavy KSVZ
  fields. While $\epsilon_{\HAD/\EM}$ are independent of $e_Q$,
  $\tau_{\stau}\propto 1/e_Q^4$. The $f_a$ limits can thus be relaxed,
  e.g., by one order of magnitude for $e_Q=1$.  
\item The CBBN and hadronic BBN constraints in the case of the
  $\selectronR$ or $\smuonR$ NLSP are identical to the ones shown.
  The electromagnetic BBN constraints however will be more restrictive
  in the $\selectronR$ NLSP case since all of the electron energy
  $E_e$ released in the $\selectronR$ NLSP decay will contribute:
  $\epsilon_{\EM}=E_e$.
\end{itemize}
%

\section{Conclusion}
\label{Sec:Conclusion}

For axino LSP scenarios with a long-lived charged slepton NLSP, we
have studied BBN constraints associated with hadronic and
electromagnetic energy release.
While the region with $f_a\lesssim 10^{12}\,\GeV$ is typically not
affected, those constraints become significant for larger $f_a$ such
that models with $f_a$ towards the grand unification scale are
disfavored.
The new BBN constraints on $f_a$ can be more restrictive than the
recently obtained CBBN constraints~\cite{Freitas:2009fb}.
This further tightens the upper limits on the reheating temperature
discussed in Ref.~\cite{Freitas:2009fb} which are relevant for models
of inflation and baryogenesis.

\begin{acknowledgments}
  N.T.\ would like to thank the Max Planck Institute for Physics for
  their kind hospitality during parts of this work.
  This research was partially supported by the Cluster of Excellence
  `Origin and Structure of the Universe' and by the Swiss National
  Science Foundation (SNF).
\end{acknowledgments}
%
\bibliographystyle{epj}


\begin{thebibliography}{48}

\bibitem{Bonometto:1993fx}
S.A. Bonometto, F.~Gabbiani, A.~Masiero, Phys. Rev. \textbf{D49}, 3918 (1994),
  \texttt{hep-ph/9305237}

\bibitem{Covi:1999ty}
L.~Covi, J.E. Kim, L.~Roszkowski, Phys. Rev. Lett. \textbf{82}, 4180 (1999),
  \texttt{hep-ph/9905212}

\bibitem{Covi:2001nw}
L.~Covi, H.B. Kim, J.E. Kim, L.~Roszkowski, JHEP \textbf{05}, 033 (2001),
  \texttt{hep-ph/0101009}

\bibitem{Covi:2004rb}
L.~Covi, L.~Roszkowski, R.~Ruiz~de Austri, M.~Small, JHEP \textbf{06}, 003
  (2004), \texttt{hep-ph/0402240}

\bibitem{Brandenburg:2004du}
A.~Brandenburg, F.D. Steffen, JCAP \textbf{0408}, 008 (2004),
  \texttt{hep-ph/0405158}

\bibitem{Steffen:2008qp}
F.D. Steffen, Eur. Phys. J. \textbf{C59}, 557 (2009), \texttt{0811.3347}

\bibitem{Baer:2008yd}
H.~Baer, M.~Haider, S.~Kraml, S.~Sekmen, H.~Summy, JCAP \textbf{0902}, 002
  (2009), \texttt{0812.2693}

\bibitem{Freitas:2009fb}
A.~Freitas, F.D. Steffen, N.~Tajuddin, D.~Wyler, Phys. Lett. \textbf{B679}, 270
  (2009), \texttt{0904.3218}

\bibitem{Choi:2007rh}
K.Y. Choi, L.~Roszkowski, R.~Ruiz~de Austri, JHEP \textbf{04}, 016 (2008),
  \texttt{0710.3349}

\bibitem{Kawasaki:2007mk}
M.~Kawasaki, K.~Nakayama, M.~Senami, JCAP \textbf{0803}, 009 (2008),
  \texttt{0711.3083}

\bibitem{Baer:2009ms}
H.~Baer, A.D. Box, H.~Summy (2009), \texttt{0906.2595}

\bibitem{Pospelov:2006sc}
M.~Pospelov, Phys. Rev. Lett. \textbf{98}, 231301 (2007),
  \texttt{hep-ph/0605215}

\bibitem{Pospelov:2007js}
M.~Pospelov (2007), \texttt{0712.0647}

\bibitem{Pospelov:2008ta}
M.~Pospelov, J.~Pradler, F.D. Steffen, JCAP \textbf{0811}, 020 (2008),
  \texttt{0807.4287}

\bibitem{Amsler:2008zz}
C.~Amsler et~al. (Particle Data Group), Phys. Lett. \textbf{B667}, 1 (2008)

\bibitem{Sikivie:2006ni}
P.~Sikivie, Lect. Notes Phys. \textbf{741}, 19 (2008),
  \texttt{astro-ph/0610440}

\bibitem{Raffelt:2006cw}
G.G. Raffelt, Lect. Notes Phys. \textbf{741}, 51 (2008),
  \texttt{hep-ph/0611350}

\bibitem{Kim:1979if}
J.E. Kim, Phys. Rev. Lett. \textbf{43}, 103 (1979)

\bibitem{Shifman:1979if}
M.A. Shifman, A.I. Vainshtein, V.I. Zakharov, Nucl. Phys. \textbf{B166}, 493
  (1980)

\bibitem{Kim:1983ia}
J.E. Kim, Phys. Lett. \textbf{B136}, 378 (1984)

\bibitem{Schilling:2005dr}
S.~Schilling, \emph{{Two-Loop Techniques in Rare Decays}} (2005), {PhD Thesis,
  University of Zuerich}

\bibitem{Brandenburg:2005he}
A.~Brandenburg, L.~Covi, K.~Hamaguchi, L.~Roszkowski, F.D. Steffen, Phys. Lett.
  \textbf{B617}, 99 (2005), \texttt{hep-ph/0501287}

\bibitem{Kawasaki:2004qu}
M.~Kawasaki, K.~Kohri, T.~Moroi, Phys. Rev. \textbf{D71}, 083502 (2005),
  \texttt{astro-ph/0408426}

\bibitem{Jedamzik:2006xz}
K.~Jedamzik, Phys. Rev. \textbf{D74}, 103509 (2006), \texttt{hep-ph/0604251}

\bibitem{Cyburt:2009pg}
R.H. Cyburt et~al. (2009), \texttt{0907.5003}

\bibitem{Reno:1987qw}
M.H. Reno, D.~Seckel, Phys. Rev. \textbf{D37}, 3441 (1988)

\bibitem{Kawasaki:2008qe}
M.~Kawasaki, K.~Kohri, T.~Moroi, A.~Yotsuyanagi, Phys. Rev. \textbf{D78},
  065011 (2008), \texttt{0804.3745}

\bibitem{Feng:2003uy}
J.L. Feng, A.~Rajaraman, F.~Takayama, Phys. Rev. \textbf{D68}, 063504 (2003),
  \texttt{hep-ph/0306024}

\bibitem{Steffen:2006hw}
F.D. Steffen, JCAP \textbf{0609}, 001 (2006), \texttt{hep-ph/0605306}

\bibitem{Kohri:2005wn}
K.~Kohri, T.~Moroi, A.~Yotsuyanagi, Phys. Rev. \textbf{D73}, 123511 (2006),
  \texttt{hep-ph/0507245}

\bibitem{Bailly:2008yy}
S.~Bailly, K.~Jedamzik, G.~Moultaka, Phys. Rev. \textbf{D80}, 063509 (2009),
\texttt{0812.0788}

\bibitem{Kim:1992eu}
J.E. Kim, Phys. Rev. Lett. \textbf{67}, 3465 (1991)

\bibitem{Lyth:1993zw}
D.H. Lyth, Phys. Rev. \textbf{D48}, 4523 (1993), \texttt{hep-ph/9306293}

\bibitem{Chang:1996ih}
S.~Chang, H.B. Kim, Phys. Rev. Lett. \textbf{77}, 591 (1996),
  \texttt{hep-ph/9604222}

\bibitem{Hashimoto:1998ua}
M.~Hashimoto, K.I. Izawa, M.~Yamaguchi, T.~Yanagida, Phys. Lett. \textbf{B437},
  44 (1998), \texttt{hep-ph/9803263}

\bibitem{Asaka:2000zh}
T.~Asaka, K.~Hamaguchi, K.~Suzuki, Phys. Lett. \textbf{B490}, 136 (2000),
  \texttt{hep-ph/0005136}

\bibitem{Ratz:2008qh}
M.~Ratz, K.~Schmidt-Hoberg, M.W. Winkler, JCAP \textbf{0810}, 026 (2008),
  \texttt{0808.0829}

\bibitem{Pradler:2008qc}
J.~Pradler, F.D. Steffen, Nucl. Phys. \textbf{B809}, 318 (2009),
  \texttt{0808.2462}

\bibitem{Pradler:2006hh}
J.~Pradler, F.D. Steffen, Phys. Lett. \textbf{B648}, 224 (2007),
  \texttt{hep-ph/0612291}

\bibitem{Spergel:2006hy}
D.N. Spergel et~al. (WMAP), Astrophys. J. Suppl. \textbf{170}, 377 (2007),
  \texttt{astro-ph/0603449}

\bibitem{Kamimura:2008fx}
M.~Kamimura, Y.~Kino, E.~Hiyama, 
  Prog.\ Theor.\ Phys.\  {\bf 121}, 1059 (2009), \texttt{0809.4772}

\bibitem{Sigl:1995kk}
G.~Sigl, K.~Jedamzik, D.N. Schramm, V.S. Berezinsky, Phys. Rev. \textbf{D52},
  6682 (1995), \texttt{astro-ph/9503094}

\bibitem{Jedamzik:1999di}
K.~Jedamzik, Phys. Rev. Lett. \textbf{84}, 3248 (2000),
  \texttt{astro-ph/9909445}

\bibitem{Jedamzik:2004er}
K.~Jedamzik, Phys. Rev. \textbf{D70}, 063524 (2004), \texttt{astro-ph/0402344}

\bibitem{Cyburt:2006uv}
R.H. Cyburt, J.R. Ellis, B.D. Fields, K.A. Olive, V.C. Spanos, JCAP
  \textbf{0611}, 014 (2006), \texttt{astro-ph/0608562}

\bibitem{Freitas:2009xxx}
A.~Freitas, F.D. Steffen, N.~Tajuddin, D.~Wyler (2009), in preparation

\bibitem{Cyburt:2002uv}
R.H. Cyburt, J.R. Ellis, B.D. Fields, K.A. Olive, Phys. Rev. \textbf{D67},
  103521 (2003), \texttt{astro-ph/0211258}

\bibitem{Asplund:2005yt}
M.~Asplund, D.L. Lambert, P.E. Nissen, F.~Primas, V.V. Smith, Astrophys. J.
  \textbf{644}, 229 (2006), \texttt{astro-ph/0510636}

\bibitem{Jedamzik:2007qk}
K.~Jedamzik, JCAP \textbf{0803}, 008 (2008), \texttt{0710.5153v2}

\bibitem{Pradler:2009dr}
J.~Pradler, \emph{{The long-lived stau as a thermal relic}} (2009), {PhD
  Thesis, Technical University Munich}, \texttt{0909.3429}

\bibitem{Mukhanov:2003xs}
V.F. Mukhanov, Int. J. Theor. Phys. \textbf{43}, 669 (2004),
  \texttt{astro-ph/0303073}

\end{thebibliography}
%
%
\end{document}